\documentclass[10pt]{article}

\usepackage{times}
\usepackage{amssymb,amsmath,amsthm}
\usepackage{mathrsfs}

\DeclareMathAlphabet\EuScript{U}{eus}{m}{n}
\SetMathAlphabet\EuScript{bold}{U}{eus}{b}{n}

\newcommand{\ie}{\emph{i.e.}}
\newcommand{\eg}{\emph{e.g.}}

\def\B{\mathscr{B}}
\def\C{\mathbb{C}}
\def\H{\mathcal H}
\def\D{\mathscr D}
\def\N{\mathbb{N}}
\def\R{\mathbb{R}}
\def\F{\mathscr F}

\def\<{\left\langle}
\def\>{\right\rangle}
\def\({\left(}
\def\){\right)}
\def\[{\left[}
\def\]{\right]}
\def\ltwo{\mathsf{L}^{\:\!\!2}}
\def\linf{\mathsf{L}^{\:\!\!\infty}}
\def\e{\mathrm{e}}
\def\de{\mathrm{d}}
\def\dom{\matheucal D}
\def\diag{\mathop{\mathrm{diag}}\nolimits}
\def\esssup{\mathop{\mathrm{ess\,sup}}}
\def\essinf{\mathop{\mathrm{ess\,inf}}}
\def\im{\mathop{\mathsf{Im}}\nolimits}
\def\re{\mathop{\mathsf{Re}}\nolimits}

\newtheorem{Theorem}{Theorem}[section]
\newtheorem{Proposition}[Theorem]{Proposition}
\newtheorem{Lemma}[Theorem]{Lemma}
\newtheorem{Remark}[Theorem]{Remark}
\newtheorem{Definition}[Theorem]{Definition}

\begin{document}
  
  
  \title{{\Large\textbf{On perturbations of Dirac operators \\
	with variable magnetic field of constant direction}}}
 \author{Serge Richard\,~and\,~Rafael Tiedra de Aldecoa}
  \date{\small
    \begin{quote}
      \emph{
    \begin{itemize}
    \item[]
      D\'epartement de Physique Th\'eorique, Universit\'e de Gen\`eve,\\
      24, quai E. Ansermet, 1211 Gen\`eve 4, Switzerland
    \item[]
      \emph{E-mails\:\!:}
      serge.richard@physics.unige.ch\,~and\,~rafael.tiedra@physics.unige.ch
    \end{itemize}
      }
    \end{quote}
    May 2004
  }
  \maketitle
  
  
  \begin{abstract}
    We carry out the spectral analysis of matrix valued perturbations of
    $3$-dimensional Dirac operators with variable magnetic field of constant
    direction. Under suitable assumptions on the magnetic field and on the
    pertubations, we obtain a limiting absorption principle, we prove the
    absence of singular continuous spectrum in certain intervals and state
    properties of the point spectrum. Various situations, for example  when
    the magnetic field is constant, periodic or diverging at infinity, are
    covered. The importance of an internal-type operator (a $2$-dimensional Dirac
    operator) is also revealed in our study. The proofs rely on commutator
    methods.
  \end{abstract}
  
  \setcounter{equation}{0}
  
  \section{Introduction and main results}\label{S1}
  
  We consider a relativistic spin-$\frac12$ particle evolving in $\R^3$ in presence
  of a variable magnetic field of constant direction. By virtue of the Maxwell
  equations, we may assume with no loss of generality that the magnetic field has
  the form $\vec B(x_1,x_2,x_3)=\big(0,0,B(x_1,x_2)\big)$. So the unperturbed
  system is described, in the Hilbert space $\ltwo(\R^3;\C^4)$, by the Dirac
  operator 
  \begin{equation*}
    H_0:=\alpha_1\Pi_1+\alpha_2\Pi_2+\alpha_3P_3+\beta m\:\!,
  \end{equation*}
  where $\beta \equiv \alpha_0, \alpha_1, \alpha_2, \alpha_3$ are the usual
  Dirac-Pauli matrices, $m$ is the strictly positive mass of the particle and
  $\Pi_j:=-i\partial_j-a_j$ are the generators of the magnetic translations with
  a vector potential $\vec a(x_1,x_2,x_3)=\(a_1(x_1,x_2),a_2(x_1,x_2),0\)$ that
  satisfies $B=\partial_1a_2-\partial_2a_1$. Since $a_3 =0$, we have written
  $P_3:=-i\partial_3$ instead of $\Pi_3$.

  In this paper we study the stability of certain parts of the spectrum of $H_0$
  under matrix valued perturbations $V$. More precisely, if $V$ satisfies some
  natural hypotheses, we shall prove the absence of singular continuous spectrum
  and the finiteness of the point spectrum of $H:=H_0+V$ in intervals of $\R$
  corresponding to gaps in the symmetrized spectrum of the operator
  $H^0:=\sigma_1\Pi_1+\sigma_2\Pi_2+\sigma_3m$ in $\ltwo(\R^2;\C^2)$. The matrices
  $\sigma_j$ are the Pauli matrices and the symmetrized spectrum
  $\sigma^0_{\rm sym}$ of $H^0$ is the union of the spectra of $H^0$ and $-H^0$.
  We stress that our analysis does not require any restriction on the behaviour
  of the magnetic field at infinity. Nevertheless, the pertinence of our work
  depends on a certain property of the internal-type operator $H^0$\:\!; 
  namely, the size and the number of gaps in $\sigma^0_{\rm sym}$. We refer to
  \cite{BS}, \cite{D}, \cite{GMo}, \cite{HNW} and \cite{Th} for various results on
  the spectrum of $H^0$, especially in the situations of physical interest, for
  example  when $B$ is constant, periodic or diverges at infinity.

  Technically, this work relies on commutator methods initiated by E.~Mourre
  \cite{MO} and extensively developed in \cite{ABG}. For brevity we shall
  constantly refer to the latter reference for notations and definitions. Our
  choice of a conjugate operator enables us to treat Dirac operators with general
  magnetic fields provided they point in a constant direction. On the other hand,
  as already put into evidence in \cite{GMa}, the use of a conjugate operator with
  a matrix structure has a few ``rather awkward consequences'' for long-range
  perturbations. We finally mention that this study is the counterpart for Dirac
  operators of \cite{MP}, where only Schr\"odinger operators are considered.
  Unfortunately, the intrinsic structure of the Dirac equation prevents us 
  from using the possible magnetic anisotropy to control
  the perturbations (see Remark \ref{Remark inequality} for details). 
  
  We give now a more precise description of our results. For simplicity we
  impose the continuity of the magnetic field and avoid perturbations with
  local singularities. Hence we assume that $B$ is a $C(\R^2;\R)$-function and
  choose any vector potential $\vec a=(a_1,a_2,0)\in C(\R^2;\R^3)$, \eg~the one
  obtained by means of the transversal gauge \cite{Th}. The definitions below
  concern the admissible perturbations. In the long-range case, we restrict them
  to the scalar type in order not to impose unsatisfactory constraints.
  In the sequel, $\mathscr B_{\rm h}(\C^4)$ stands for the set of $4 \times 4$ 
  hermitian matrices, and $\|\cdot\|$ denotes the norm of the Hilbert space 
  $\H:=\ltwo(\R^3;\C^4)$ as well as the norm of $\B(\H)$, the set of bounded 
  linear operators on $\H$. $\N:=\{0,1,2,\ldots\}$ is the set of natural numbers.
  $\vartheta$ is an arbitrary $C^\infty\!\big([0,\infty)\big)$-function such that
  $\vartheta=0$ near $0$ and $\vartheta=1$ near infinity. $Q_j$ is the multiplication
  operator by the coordinate $x_j$ in $\H$, and the expression
  $\langle\:\!\cdot\:\!\rangle$ corresponds to $\sqrt{1+(\:\!\cdot\:\!)^2}$.
    
  \begin{Definition}\label{defdespots}
    Let $V$ be a multiplication operator associated with an element of
    \:\!$\linf\big(\R^3;\mathscr B_{\rm h}(\C^4)\big)$.
    \begin{itemize}
    \item[(a)] $V$ is \:\!\emph{small at infinity}\:\! if~
      $\displaystyle\lim_{r\to\infty}\Big\|
      \vartheta\Big(\frac{\langle Q\rangle}r\Big)V\Big\|=0$\:\!,
    \item[(b)] $V$ is \:\!\emph{short-range}\:\! if~
      $\displaystyle\int_1^\infty\Big\|\vartheta
      \Big(\frac{\langle Q_3\rangle}r\Big)V\Big\|\,\de r <\infty$\:\!,
    \item[(c)] Let $V_{\scriptscriptstyle\textrm L}$ be in
      $C^1(\R^3;\R)$ with
      $x\mapsto\langle x_3\rangle(\partial_j V_{\scriptscriptstyle\textrm L})(x)$
      in $\linf(\R^3;\R)$ for $j=1,2,3,$ then $V:=V_{\scriptscriptstyle\textrm L}$ is
      \:\!\emph{long-range}\:\! if
      \begin{equation*}
	\int_1^{\infty}\Big\|\vartheta
	\Big(\frac{\langle Q_3 \rangle}r\Big)\langle Q_3\rangle
	\:\!(\partial_j V)\Big\|\,\frac{\de r}r <\infty\qquad
	\textrm{for}\qquad j=1,2,3\:\!.
      \end{equation*}
    \end{itemize}
  \end{Definition}
  
  Note that Definitions \ref{defdespots}.(b) and \ref{defdespots}.(c) differ from
  the standard ones\:\!: the decay rate is imposed only in the $x_3$ direction.

  We are in a position to state our results. Let $\dom(\langle Q_3 \rangle)$ denote
  the domain of $\langle Q_3\rangle$ in $\H$, then the limiting absorption principle for
  $H$ is expressed in terms of the Banach space
  $\mathcal G:=\big(\dom(\langle Q_3\rangle),\H\big)_{1/2,1}$ defined by real interpolation
  \cite{ABG}. For convenience, we recall that $\dom(\langle Q_3\rangle^s)$ is contained
  in $\mathcal G$ for each $s>1/2$.
  
  \begin{Theorem}\label{final thm} 
    Assume that $B$ belongs to $C(\R^2;\R)$, and that $V$ belongs to
    $\linf\big(\R^3;\mathscr B_{\rm h}(\C^4)\big)$, is small at infinity and can be written
    as the sum of a short-range and a long-range matrix valued function. Then
    \begin{itemize}
    \item[(a)] The point spectrum of the operator $H$ in
      \:\!$\R\setminus\sigma^0_{\rm sym}$\:\! is composed of eigenvalues of finite
      multiplicity and with no accumulation point in
      \:\!$\R\setminus\sigma^0_{\rm sym}$\:\!. 
    \item[(b)] The operator $H$ has no singular continuous spectrum in
      \:\!$\R\setminus \sigma^0_{\rm sym}$\:\!.
    \item[(c)] The limits
      $\lim_{\varepsilon\to+0}\langle\psi,(H-\lambda\mp i\varepsilon)^{-1}\psi\rangle$ exist
      for each $\psi\in\mathcal G$, uniformly in $\lambda$ on each compact subset of
      \:\!$\R\setminus\{\sigma_{\rm sym}^0\cup\sigma_{\rm pp}(H)\}$. 
    \end{itemize}  
  \end{Theorem}
  
  The above statements seem to be new for such a general magnetic field. In the special
  but important case of a nonzero constant magnetic field $B_0$, the admissible
  perturbations introduced in Definition \ref{defdespots} are more general than those
  allowed in \cite{Y}. We stress that in this situation $\sigma_{\rm sym}^0$ is equal
  to $\{\pm\sqrt{2nB_0+m^2}\,:\,n\in\N\}$, which implies that there are plenty of gaps
  where our analysis gives results. On the other hand, if $B(x_1,x_2)\to0$ as
  $|(x_1,x_2)|\to\infty$, our treatment gives no information since both $(-\infty,-m]$
  and $[m,\infty)$ belong to $\sigma^0_{\rm sym}$. We finally mention the paper \cite{BC}
  for a related work on perturbations of magnetic Dirac operators.
  
  \section{Mourre estimate for the operator $H_0$}
  
  \subsection{Preliminaries}
 
  Let us start by recalling some known results. The operator $H_0$ is essentially
  self-adjoint on $\D:=C_0^\infty(\R^3;\C^4)$ \cite[Thm.~2.1]{Che}. Its spectrum is
  symmetric with respect to $0$ and does not contain the interval $(-m,m)$
  \cite[Cor.~5.14]{Th}. Thus the subset $H_0\D$ is dense in $\H$ since $\D$ is dense
  in $\dom(H_0)$ (endowed with the graph topology) and $H_0$ is a homeomorphism from
  $\dom(H_0)$ onto $\H$.

  We now introduce a suitable representation of the Hilbert space $\H$. We consider
  the partial Fourier transformation
  \begin{equation}\label{F unitaire}
    \F:\D\to\int_\R^\oplus\H_{\scriptscriptstyle 12}\,\de\xi\:\!,
    \qquad(\F\psi)(\xi):=\frac1{\sqrt{2\pi}}
    \int_\R\e^{-i\xi x_3}\psi(\:\!\cdot\:\!,x_3)\,\de x_3\:\!,
  \end{equation}
  where $\H_{\scriptscriptstyle 12}:=\ltwo(\R^2;\C^4)$. This map extends uniquely to
  a unitary operator from $\H$ onto $\int_\R^\oplus\H_{\scriptscriptstyle 12}\,\de\xi$,
  which we denote by the same symbol $\mathscr F$. As a first application, one obtains
  the following direct integral decomposition of $H_0$\:\!:
  \begin{equation*}
    \F H_0\F^{-1}=\int_\R^\oplus H_0(\xi)\,\de\xi\:\!,
  \end{equation*}
  where $H_0(\xi)$ is a self-adjoint operator in $\H_{\scriptscriptstyle 12}$ acting
  as $\alpha_1\Pi_1+\alpha_2\Pi_2+\alpha_3\xi+\beta m$ on $C_0^\infty(\R^2;\C^4)$.
  In the following remark we draw the connection between the operator $H_0(\xi)$ and
  the operator $H^0$ introduced in Section \ref{S1}. It reveals the importance of the
  internal-type operator $H^0$ and shows why its negative $-H_0$ also has to be taken
  into account. 
  
  \begin{Remark}
    The operator $H_0(0)$ acting on $C_0^{\infty}(\R^2;\C^4)$ is unitarily equivalent
    to the direct sum operator
    $
    \footnotesize
    \begin{pmatrix}
      m & \Pi_-\\
      \Pi_+ & -m
    \end{pmatrix}
    \oplus
    \begin{pmatrix}
      m & \Pi_+
      \\ \Pi_- & -m
    \end{pmatrix}
    $
    acting on $C_0^\infty(\R^2;\C^2)\oplus C_0^\infty(\R^2;\C^2)$, where
    $\Pi_\pm:=\Pi_1\pm i\Pi_2$. Now, these two matrix operators act in $\ltwo(\R^2;\C^2)$
    and are essentially self-adjoint on $C_0^\infty(\R^2; \C^2)$ \cite[Thm.~2.1]{Che}.
    However, the first one is nothing but $H^0$, while the second one is unitarily
    equivalent to $-H^0$ (this can be obtained by means of the abstract Foldy-Wouthuysen
    transformation \cite[Thm.~5.13]{Th}). Therefore $H_0(0)$ is essentially self-adjoint
    on $C_0^{\infty}(\R^2;\C^4)$ and
    \begin{equation*}
      \sigma[H_0(0)]=\sigma(H^0)\cup\sigma(-H^0)
      \equiv\sigma^0_{\rm sym}\:\!.
    \end{equation*} 
    Moreover, there exists a relation between $\sigma[H_0(\xi)]$ and $\sigma^0_{\rm sym}$.
    Indeed, for $\xi\in\R$ fixed, one can show that $H_0(\xi)^2=H_0(0)^2+\xi^2$ on
    $\dom\big(H_0(\xi)^2\big)=\dom\big(H_0(0)^2\big)$, so that
    \begin{equation}
      \sigma[H_0(\xi)^2]=\sigma[H_0(0)^2+\xi^2]=\big(\sigma[H_0(0)]\big)^2+\xi^2
      =(\sigma^0_{\rm sym})^2+\xi^2\:\!,\label{peuplus}
    \end{equation}
    where the spectral theorem has been used for the second equality. Since the spectrum
    of $H_0(\xi)$ is symmetric with respect to $0$ \cite[Cor.~5.14]{Th}, it follows that 
    \begin{equation*}
      \sigma[H_0(\xi)]=-\sqrt{(\sigma^0_{\rm sym})^2+\xi^2}
      \,\cup\,\sqrt{(\sigma^0_{\rm sym})^2+\xi^2}\:\!.
    \end{equation*}
    Define $\mu_0:=\inf|\sigma^0_{\rm sym}|$ (which is bigger or equal to $m$ because
    $H^0$ has no spectrum in $(-m,m)$ \cite[Cor.~5.14]{Th}). Then from the direct
    integral decomposition of $H_0$, one readily gets
    \begin{equation}\label{spectreH_0}
      \sigma(H_0)=(-\infty,-\mu_0]\cup[\mu_0,+\infty)\:\!.
    \end{equation}
  \end{Remark}
  
  We conclude the section by giving two technical lemmas in relation with the operator
  $H_0^{-1}$. Proofs can be found in an appendix.

  \begin{Lemma}\label{P_3H_0^{-1}}
    \begin{itemize}
    \item[(a)] For each $n\in \N$, $H_0^{-n}\D$ belongs to $\dom(Q_3)$,
    \item[(b)] $P_3H_0^{-1}$ is a bounded self-adjoint operator equal to
      $H_0^{-1}P_3$ on $\dom(P_3)$. In particular, $H_0^{-1}\H$ belongs to $\dom(P_3)$.
  \end{itemize}
  \end{Lemma}
  
  One may observe that, given a $C^1(\R;\C)$-function $f$ with $f'$ bounded, the
  operator $f(Q_3)$ is well-defined on $\dom(Q_3)$. Thus $f(Q_3)H_0^{-n}\D$ is a
  subset of $\H$ for each $n\in\N$. The preceding lemma and the following simple
  statement are constantly used in the sequel. 

  \begin{Lemma}\label{Hetf}
    Let $f$ be in $C^1(\R;\C)$ with $f'$ bounded, and $n\in\N$. Then
    \begin{itemize}
    \item[(a)]  
      $iH_0^{-1}f(Q_3)-if(Q_3)H_0^{-1}$ is equal to $-H_0^{-1}\alpha_3f'(Q_3)H_0^{-1}$
      on $H_0^{-n}\D$,
    \item[(b)] $P_3H_0^{-1}f(Q_3)-f(Q_3)P_3H_0^{-1}$ is equal to
      $i(P_3H_0^{-1}\alpha_3-1)f'(Q_3)H_0^{-1}$ on $\D$.
    \end{itemize}
  \end{Lemma}
  
  Both right terms belong to $\B(\H)$. For shortness we shall denote them by
  $[iH_0^{-1},f(Q_3)]$ and $[P_3H_0^{-1},f(Q_3)]$ respectively.
  
  \subsection{The conjugate operator}
  
  The aim of the present section is to define an appropriate operator conjugate
  to $H_0$. To begin with, one observes that $Q_3P_3H_0^{-1}\D\subset\H$ as a
  consequence of Lemma \ref{P_3H_0^{-1}}. In particular, the formal expression
  \begin{equation}\label{defdeA}
    A:=\mbox{$\frac12$}(H_0^{-1}P_3Q_3+Q_3P_3H_0^{-1})
  \end{equation}
  leads to a well-defined symmetric operator on $\D$.

  \begin{Proposition}\label{sur A}
    The operator $A$ is essentially self-adjoint on $\D$ and its closure is
    essentially self-adjoint on any core for $\langle Q_3\rangle$.
  \end{Proposition}

  \begin{proof}
    The claim is a consequence of Nelson's criterion of essential self-adjointness
    \cite[Thm.~X.37]{RS} applied to the triple $\{\langle
    Q_3\rangle,A,\D\}$. Let us simply verify the two hypotheses of that theorem. 
    By using Lemmas
    \ref{P_3H_0^{-1}} and \ref{Hetf}, one first obtains that for all
    $\psi\in\D$\:\!:
    \begin{equation*}
      \|A\psi\|=\left\|\(P_3H_0^{-1}Q_3-\mbox{$\frac12$}
      \[P_3H_0^{-1}, Q_3\]\)\psi\right\|\leq\textsc c\;\!\|\langle Q_3\rangle\psi\|
    \end{equation*}
    for some constant $\textsc c>0$ independent of $\psi$. Then, for all $\psi\in\D$
    one has\:\!:
    \begin{align*}
      \langle A\psi,\langle Q_3\rangle\psi\rangle
      -\langle\langle Q_3\rangle\psi,A\psi\rangle
      &=i\im\< Q_3\psi,\[P_3H_0^{-1}
      ,\langle Q_3\rangle\]\psi\>\\
      &=i\re\<\(\alpha_3P_3H_0^{-1}-1\)Q_3\psi,
      Q_3\langle Q_3\rangle^{-1}H_0^{-1}\psi\>.
    \end{align*}
    A few more commutator calculations, using again Lemma \ref{Hetf} with
    $f(Q_3)=\langle Q_3\rangle^{1/2}$, lead to the following result\:\!: for all
    $\psi\in\D$, there exists a constant $\textsc d>0$ independent of $\psi$ such
    that
    \begin{equation*}
      |\<A\psi,\<Q_3\>\psi\>-\<\<Q_3\>\psi,A\psi\>|
      \leq\textsc d\;\!\big\|\<Q_3\>^{\frac12}\psi\big\|^2.\qedhere
    \end{equation*}
  \end{proof}
  
  As far as we know, the matrix conjugate operator (\ref{defdeA}) has never been
  employed before for the study of magnetic Dirac operators. 
  
  \subsection{Strict Mourre estimate for $H_0$}
  
  We now gather some results on the regularity of $H_0$ with respect to $A$. We
  recall that $\dom(H_0)^*$ is the adjoint space of $\dom(H_0)$ and that one has
  the continuous dense embeddings $\dom(H_0)\hookrightarrow\H\hookrightarrow\dom(H_0)^*$,
  where $\H$ is identified with its adjoint through the Riesz isomorphism. 

  \begin {Proposition}\label{H_0 regularity}
    \begin{itemize}
    \item[(a)] The quadratic form
      $\dom(A)\ni\psi\mapsto\langle H_0^{-1}\psi,iA\psi\rangle
      -\langle A\psi,iH_0^{-1}\psi\rangle$ extends uniquely to the bounded 
      form defined by the operator $-H_0^{-1}(P_3H_0^{-1})^2H_0^{-1}\in\B(\H)$.
    \item[(b)] The group $\{e^{itA}\}_{t\in\R}$ leaves $\dom(H_0)$
      invariant.
    \item[(c)] The quadratic form
      \begin{equation}\label{form1}
	\dom(A)\ni\psi\mapsto\langle H_0^{-1}(P_3H_0^{-1})^2H_0^{-1}\psi,
	iA\psi\rangle-\langle A\psi,iH_0^{-1}(P_3H_0^{-1})^2H_0^{-1}\psi\rangle,
      \end{equation} 
      extends uniquely to a bounded form on $\H$.
    \end{itemize}
  \end{Proposition}

  In the framework of \cite{ABG}, the statements of (a) and (c) mean
  that $H_0$ is of class $C^1(A)$ and $C^2(A)$ respectively.

  \begin{proof}
    (a) For any $\psi\in\D$, one gets
    \begin{align}\label{f1}
      \nonumber
      2\(\langle H_0^{-1}\psi,iA\psi\rangle
      -\langle A\psi,iH_0^{-1}\psi\rangle\)
      &=\langle[iH_0^{-1},Q_3]\psi,P_3H_0^{-1}\psi\rangle
      +\langle P_3H_0^{-1}\psi,[iH_0^{-1},Q_3]\psi\rangle\\
      &=-\langle H_0^{-1}\psi,(\alpha_3P_3H_0^{-1}
      +H_0^{-1}\alpha_3P_3)H_0^{-1}\psi\rangle,
    \end{align}
    where we have used Lemmas \ref{P_3H_0^{-1}} and \ref{Hetf} . Furthermore, one has
    \begin{equation}\label{f2}
      H_0^{-1}\alpha_3=-\alpha_3H_0^{-1}+2H_0^{-1}P_3H_0^{-1}
    \end{equation}
    as an operator identity in $\B(\H)$. When inserting (\ref{f2}) into (\ref{f1}),
    one obtains the equality
    \begin{equation}\label{f3}
      \langle H_0^{-1}\psi,iA\psi\rangle-\langle A \psi,iH_0^{-1}\psi\rangle
      =-\langle\psi,H_0^{-1}(P_3 H_0^{-1})^2H_0^{-1}\psi\rangle.  
    \end{equation}
    Since $\D$ is a core for $A$, the statement is obtained by density. We shall
    write $[iH_0^{-1}, A]$ for the bounded extension of the quadratic form
    $\dom(A)\ni\psi\mapsto\langle H_0^{-1}\psi,iA\psi\rangle
    -\langle A\psi,iH_0^{-1}\psi\rangle$.

    (b) Since $\dom(H_0)$ is not explicitly known, one has to invoke an abstract
    result in order to show the invariance. 
    Let $[iH_0,A]$ be the operator in $\B\(\dom(H_0), \dom(H_0)^*\)$
    associated with the unique extension to $\dom(H_0)$ of the quadratic form
    $\psi\mapsto\langle H_0\psi,iA\psi\rangle-\langle A\psi,iH_0\psi\rangle$ defined
    for all $\psi\in\dom(H_0)\cap\dom(A)$.
    Then $\dom(H_0)$ is invariant under $\{e^{itA}\}_{t \in \R}$ if $H_0$ is
    of class $C^1(A)$ and if $[iH_0,A]\dom(H_0)\subset\H$ \cite[Lemma~2]{GG}.
    From equation (\ref{f3}) and \cite[Eq.~6.2.24]{ABG}, one obtains the
    following equalities valid in form sense on $\H$\:\!:
    \begin{equation*}
      -H_0^{-1}(P_3H_0^{-1})^2H_0^{-1}=\[iH_0^{-1},A\]=-H_0^{-1}\[iH_0,A\]H_0^{-1}.
    \end{equation*}
    Thus $\[iH_0,A\]$ and $(P_3H_0^{-1})^2$ are equal as operators in 
    $\B\(\dom(H_0),\dom(H_0)^*\)$. But since the latter belongs to $\B(\H)$,
    $[iH,A]\dom(H_0)$ is included in $\H$.

    (c) The boundedness on $\D$ of the quadratic form (\ref{form1}) follows by
    inserting (\ref{defdeA}) into the r.h.s.~term of (\ref{form1}) and by applying
    repeatedly Lemma \ref{Hetf} with $f(Q_3)=Q_3$. Then one concludes by using
    the density of $\D$ in $\dom(A)$.
  \end{proof}

  From now on we shall simply denote the closure in $\H$ of $[iH_0,A]$ by 
  $T=(P_3H_0^{-1})^2\in\B(\H)$. One interest of this operator is that $\F T\F^{-1}$
  is boundedly decomposable \cite[Prop.~3.6]{Cho}, more precisely\:\!:
  \begin{equation*}
    \F T\F^{-1}=\int_\R^\oplus T(\xi)\,\de\xi\qquad\textrm{with}\qquad 
    T(\xi)=\xi^2 H_0(\xi)^{-2}\in\B(\H_{\scriptscriptstyle 12}).
  \end{equation*} 

  In the following definition, we introduce two functions giving the optimal
  value to a Mourre-type inequality. Remark that slight modifications have been
  done with regard to the usual definition \cite[Sec.~7.2.1]{ABG}.

  \begin{Definition}
    Let $H$ be a self-adjoint operator in a Hilbert space $\H$ and assume 
    that $S$ is a symmetric operator in $\B\big(\dom(H),\dom(H)^*\big)$. Let
    $E^H(\lambda;\varepsilon):= E^H\big((\lambda-\varepsilon,\lambda+
    \varepsilon)\big)$ be the spectral projection of $H$ for the interval
    $(\lambda-\varepsilon,\lambda+\varepsilon)$. Then, for all
    $\lambda\in\R$\:\! and $\varepsilon>0$, we set
    \begin{align*}
      \varrho^S_H(\lambda;\varepsilon)&:=\sup\left\{a\in\R\,:\,
      E^H(\lambda;\varepsilon)\:\!SE^H(\lambda;\varepsilon)
      \geq a\:\!E^H(\lambda;\varepsilon)\right\},\\
      \varrho^S_H(\lambda)&:=\sup_{\varepsilon>0}
      \varrho^S_H(\lambda;\varepsilon)\:\!.
    \end{align*}
  \end{Definition}

  Let us make three observations\:\!: the inequality
  $\varrho_H^S(\lambda;\varepsilon')\leq\varrho_H^S(\lambda;\varepsilon)$ holds
  whenever $\varepsilon'\geq\varepsilon$, $\varrho^S_H(\lambda)=+\infty$ if $\lambda$ 
  does not belong to the spectrum of $H$, and $\varrho^S_H(\lambda)\geq0$ for all
  $\lambda\in\R$ if $S\geq0$. We also mention that in the case of two self-adjoint
  operators $H$ and $A$ in $\H$, with $H$ of class $C^1(A)$ and $S:=\[iH,A\]$, the
  function $\varrho_H^S(\:\!\cdot\:\!)$ is equal to the function
  $\varrho_H^A(\:\!\cdot\:\!)$ defined in \cite[Eq.~7.2.4]{ABG}.

  Taking advantage of the direct integral decomposition of $H_0$ and $T$, one
  obtains for all $\lambda\in\R$ and $\varepsilon >0$\:\!:
  \begin{equation}\label{unmercredimatin}
    \varrho^T_{H_0}(\lambda; \varepsilon)=\essinf_{\xi\in\R}
    \varrho^{T(\xi)}_{H_0(\xi)}(\lambda;\varepsilon)\:\!.
  \end{equation}
  Now we can deduce a lower bound for $\varrho_{H_0}^T(\:\!\cdot\:\!)$.
  
  \begin{Proposition}\label{Mourre for H_0}
    One has
    \begin{equation}\label{unlundimatin}
      \varrho_{H_0}^T(\lambda)\geq\inf\Big\{
      \frac{\lambda^2-\mu^2}{\lambda^2}~:~\mu\in
      \sigma_{\rm sym}^0\cap\[0,|\lambda|\]\Big\} 
    \end{equation}
    with the convention that the infimum over an empty set is $+\infty$.
  \end{Proposition}
  
  \begin{proof}
    We first consider the case $\lambda\geq0$.
    
    (i) Recall from (\ref{spectreH_0}) that $\mu_0\equiv\inf|\sigma^0_{\rm sym}|=
    \inf\{\sigma(H_0)\cap [0,+\infty)\}$. Thus, for $\lambda\in[0,\mu_0)$ the l.h.s.~term
    of (\ref{unlundimatin}) is equal to $+\infty$, since $\lambda$ does not belong to the
    spectrum of $H_0$. Hence (\ref{unlundimatin}) is satisfied on $[0,\mu_0)$.

    (ii) If $\lambda\in\sigma^0_{\rm sym}$, then the r.h.s.~term of (\ref{unlundimatin}) is
    equal to $0$. However, since $T$ is positive, ${\varrho_{H_0}^T(\lambda)\geq0}$.
    Hence the relation (\ref{unlundimatin}) is again satisfied. 

    (iii) Let $0<\varepsilon<\mu_0<\lambda$. Direct computations using the explicit form
    of $T(\xi)$ and the spectral theorem for the operator $H_0(\xi)$ show that for $\xi$
    fixed, one has
    \begin{equation}\label{verre}
      \varrho^{T(\xi)}_{H_0(\xi)}(\lambda;\varepsilon)
      =\inf\Big\{\frac{\xi^2}{\rho^2}\,:\,\rho\in
      (\lambda-\varepsilon,\lambda+\varepsilon)\cap\sigma[H_0(\xi)]\Big\}
      \geq\frac{\xi^2}{(\lambda+\varepsilon)^2}\:\!.
    \end{equation}
    On the other hand one has $\varrho^{T(\xi)}_{H_0(\xi)}(\lambda;\varepsilon)
    =+\infty$ if
    $(\lambda-\varepsilon,\lambda+\varepsilon)\cap\sigma[H_0(\xi)]=\varnothing$, and a
    fortiori
    \begin{equation*}
      \varrho^{T(\xi)}_{H_0(\xi)}(\lambda;\varepsilon)=+\infty\qquad\textrm{if}\qquad
      \big((\lambda-\varepsilon)^2,(\lambda+\varepsilon)^2\big)\cap
      \sigma[H_0(\xi)^2]=\varnothing\:\!.
    \end{equation*}
    Thus, by taking into account equation (\ref{unmercredimatin}), (\ref{verre}), the
    previous observation and relation (\ref{peuplus}), one obtains that
    \begin{equation}\label{estimation1}
      \varrho^T_{H_0}(\lambda;\varepsilon)\geq\essinf
      \Big\{\frac{\xi^2}{(\lambda+\varepsilon)^2}\,:\,\xi^2\in
      \big((\lambda-\varepsilon)^2,(\lambda+\varepsilon)^2\big)
      -(\sigma^0_{\rm sym})^2\Big\}.
    \end{equation}
    Suppose now that $\lambda\not\in\sigma^0_{\rm sym}$, define
    $\mu:=\sup\{\sigma^0_{\rm sym}\cap[0,\lambda]\}$ and choose $\varepsilon>0$ such that
    $\mu<\lambda-\varepsilon$. Then the inequality (\ref{estimation1}) implies that
    \begin{equation*}
      \varrho^T_{H_0}(\lambda;\varepsilon)\geq\frac{(\lambda-\varepsilon)^2-\mu^2}
	     {(\lambda+\varepsilon)^2}\:\!.
    \end{equation*}
    Hence the relation (\ref{unlundimatin}) follows from the above formula when
    $\varepsilon\to0$.

    For $\lambda<0$, similar arguments lead to the inequality
    \begin{equation*}
      \varrho_{H_0}^T(\lambda)\geq\inf\Big\{\frac{\lambda^2-\mu^2}{\lambda^2}
      \,:\,\mu\in\sigma^0_{\rm sym}\cap\[\lambda,0\]\Big\}.
    \end{equation*} 
    The claim is then a direct consequence of the symmetry of $\sigma_{\rm sym}^0$ with
    respect to $0$.
  \end{proof}
  
  The above proposition implies that we have a strict Mourre estimate, 
  \ie~$\varrho_{H_0}^T(\:\!\cdot\:\!)>0$, on $\R\setminus\sigma_{\rm
  sym}^0$. Moreover it is not difficult to prove that $\varrho_{H_0}^T(\lambda)=0$ whenever
  $\lambda\in\sigma_{\rm sym}^0$. It follows that the conjugate operator $A$ does not
  allow to get spectral informations on $H_0$ in the subset $\sigma_{\rm sym}^0$.

  \section{Mourre estimate for the perturbed Hamiltonian}
  
  In the sequel, we consider the self-adjoint operator $H:=H_0+V$ with a
  potential $V$ that belongs to $\linf\big(\R^3;\mathscr B_{\rm h}(\C^4)\big)$.
  The domain of $H$ is equal to the domain $\dom(H_0)$ of $H_0$.   
  We first give a result on the difference of the resolvents 
  $(H-z)^{-1}-(H_0 -z)^{-1}$ and, as a corollary, we obtain the localization 
  of the essential spectrum of $H$.

  \begin{Proposition}\label{compactness}
    Assume that $V$ is small at infinity. Then for all 
    $z\in\C\setminus\big(\sigma(H)\cup\sigma(H_0)\big)$
    the difference $(H-z)^{-1}-(H_0-z)^{-1}$ is a compact operator. 
    It follows in particular that 
    $\sigma_\mathrm{ess}(H)=\sigma_\mathrm{ess}(H_0)$.
  \end{Proposition}
  
  \begin{proof}
    Since $V$ is bounded and small at infinity, it is enough to check that $H_0$
    is locally compact \cite[Sec.~4.3.4]{Th}. However, the continuity of $\vec a$
    implies that $\dom(H_0)\subset\H^{1/2}_{\rm loc}$ \cite[Thm.~1.3]{BP}. Hence
    the statement follows by usual arguments.
  \end{proof}

  \begin{Remark}\label{Remark inequality}
    In the study of an analogous problem for Schr\"odinger operators \cite{MP},
    the authors prove a result similar to Proposition \ref{compactness} without
    assuming that the perturbation is small at infinity (it only has to be small
    with respect to $B$ in a suitable sense). Their proof mainly relies on the
    structural inequalities $H_{\rm Sch}:=\Pi_1^2+\Pi_2^2+P_3^2\geq\pm B$. In the
    Dirac case, the counterpart of these turn out to be
    \begin{equation*}
      H_0^2\geq2B\cdot\diag(0,1,0,1)\qquad\textrm{and}\qquad
      H_0^2\geq-2B\cdot\diag(1,0,1,0),
    \end{equation*}
    where $\diag(\cdots)$ stands for a diagonal matrix. If we assume that the
    magnetic field is bounded from below, the first inequality enables us to treat
    pertubations of the type $\diag(V_1,V_2,V_3,V_4)$ with $V_2$, $V_4$ small with
    respect to the magnetic field and $V_1$, $V_3$ small at infinity in the original
    sense. If the magnetic field is bounded from above, the second inequality
    has to be used and the role of $V_2$, $V_4$ and $V_1$, $V_3$ are interchanged.
    However the unnatural character of these perturbations motivated us not to include
    their treatment in this paper.
  \end{Remark}
 
  In order to obtain a limiting absorption principle for $H$, one has to invoke some
  abstract results. An optimal regularity condition of $H$ with respect to $A$ has
  to be satisfied. We refer to \cite[Chap.~5]{ABG} for the definitions of
  $\mathscr C^{1,1}(A)$ and $\mathscr C^{1,1}(A;\dom(H_0),\dom(H_0)^*)$, and for more
  explanations on regularity conditions.

  \begin{Proposition}\label{V regularity}
    Let $V$ be a short-range or a long-range potential. Then $H$ is of class
    $\mathscr C^{1,1}(A)$.
  \end{Proposition}

  \begin{proof}
    Since $\{e^{itA}\}_{t\in\R}$ leaves $\dom(H)=\dom(H_0)$ invariant, it is equivalent
    to prove that $H$ belongs to $\mathscr C^{1,1}(A;\dom(H_0),\dom(H_0)^*)$
    \cite[Thm.~6.3.4.(b)]{ABG}. But in Proposition \ref{H_0 regularity}.(c), it has already
    been shown that $H_0$ is of class $C^2(A)$, so that $H_0$ is of class
    $\mathscr C^{1,1}(A;\dom(H_0),\dom(H_0)^*)$. Thus it is enough to prove that $V$ belongs to
    $\mathscr C^{1,1}(A;\dom(H_0),\dom(H_0)^*)$. In the short-range case, we shall use
    \cite[Thm.~7.5.8]{ABG}, which implies that $V$ belongs to
    $\mathscr C^{1,1}(A;\dom(H_0),\dom(H_0)^*)$. The conditions needed for that theorem
    are obtained in points (i) and (ii) below. In the long-range case, the claim follows
    by \cite[Thm.~7.5.7]{ABG}, which can be applied because of points (i), (iii), (iv)
    and (v) below. 

    (i) We first check that $\{e^{it\langle Q_3\rangle}\}_{t\in\R}$ is a polynomially
    bounded $C_0$-group in $\dom(H_0)$ and in $\dom(H_0)^*$. Lemma \ref{Hetf}.(a) (with
    $n=0$ and $f(Q_3)=\langle Q_3\rangle$) implies that $H_0$ is of class
    $C^1(\langle Q_3\rangle)$. Furthermore, by an argument similar to that given in
    part (b) of the proof of Proposition \ref{H_0 regularity}, one shows that
    $\{e^{it\langle Q_3\rangle}\}_{t\in\R}$ leaves $\dom(H_0)$ invariant. Since
    $H_0e^{it\langle Q_3\rangle}-e^{it\langle Q_3\rangle}H_0$, defined on $\D$,
    extends continuously to the operator
    $t\alpha_3Q_3\langle Q_3\rangle^{-1}e^{it\langle Q_3\rangle}\in\B(\H)$, one gets
    that
    $\|e^{it\langle Q_3\rangle}\|_{\B\(\dom(H_0)\)}\leq\textrm{Const.}\langle t\rangle$
    for all $t\in\R$, \ie the polynomial bound of the $C_0$-group in $\dom(H_0)$. By
    duality, $\{e^{it\langle Q_3\rangle}\}_{t\in\R}$ extends to a polynomially bounded
    $C_0$-group in $\dom(H_0)^*$ \cite[Prop.~6.3.1]{ABG}. 
    The generators of these $C_0$-groups are densely defined and closed in 
    $\dom(H_0)$ and in $\dom(H_0)^*$ respectively; both are simply denoted by 
    $\langle Q_3\rangle$.

    (ii) Since $\{e^{itA}\}_{t\in\R}$ leaves $\dom(H_0)$ invariant, one may also consider 
    the $C_0$-group in $\dom(H_0)$ obtained by restriction and the $C_0$-group in
    $\dom(H_0)^*$ obtained by extension. The generator of each of these $C_0$-groups will be
    denoted by $A$. Let
    $\dom\big(A;\dom(H_0)\big):=\{\varphi\in\dom(H_0)\cap\dom(A)\,:\,A\varphi\in\dom(H_0)\}$
    be the domain of $A$ in $\dom(H_0)$, and let
    $\dom\big(A^2;\dom(H_0)\big):=\{\varphi\in\dom(H_0)\cap\dom(A^2)
    \,:\,A\varphi,A^2\varphi\in\dom(H_0)\}$ be the domain of $A^2$ in $\dom(H_0)$. We now
    check that $\langle Q_3\rangle^{-1}A$ and $\langle Q_3\rangle^{-2}A^2$, defined
    on $\dom\big(A;\dom(H_0)\big)$ and on $\dom\big(A^2;\dom(H_0)\big)$ respectively, extend
    to operators in $\B\big(\dom(H_0)\big)$. After some commutator calculations performed on
    $\D$ and involving Lemma \ref{Hetf}, one first obtains that  $\langle Q_3\rangle^{-1}A$
    and $\langle Q_3\rangle^{-2}A$ are respectively equal on $\D$ to some operators $S_1$
    and $S_2\langle Q_3\rangle^{-1}$ in $\B(\H)$, where $S_1$ and $S_2$ are polynomials in
    $H_0^{-1}$, $P_3H_0^{-1}$, $\alpha_3$ and $f(Q_3)$ for bounded functions $f$ with
    bounded derivatives. Since $\D$ is a core for $A$, these equalities even hold on
    $\dom(A)$. Hence one has on $\dom(A^2)$\:\!:
    \begin{equation*}
      \langle Q_3\rangle^{-2}A^2=\(\langle Q_3\rangle^{-2}A\)A
      =S_2\langle Q_3\rangle^{-1}A=S_2S_1\:\!.
    \end{equation*}
    In consequence, $\langle Q_3\rangle^{-1}A$ and $\langle Q_3\rangle^{-2}A^2$ are equal
    on $\dom(A)$ and on $\dom(A^2)$ respectively, to operators expressed only in terms of
    $H_0^{-1}$, $P_3H_0^{-1}$, $\alpha_3$ and $f(Q_3)$ for bounded functions $f$ with
    bounded derivatives. Moreover, one easily observes that these operators and their
    products belong to $\B\big(\dom(H_0)\big)$. Thus, it follows that
    $\langle Q_3\rangle^{-1}A$ and $\langle Q_3\rangle^{-2}A^2$ are equal on
    $\dom\big(A;\dom(H_0)\big)$ and on $\dom\big(A^2;\dom(H_0)\big)$ respectively to some
    operators belonging to $\B\big(\dom(H_0)\big)$.

    (iii) By duality, the operator $(\langle Q_3\rangle^{-1}A)^*$ belongs to 
    $\B\big(\dom(H_0)^*\big)$. Now, for $\psi\in\dom(H_0)^*$ and
    $\varphi\in\dom\big(A;\dom(H_0)\big)$, one has
    \begin{equation}\label{cidessus}
      \langle(\langle Q_3\rangle^{-1}A)^*\psi,\varphi\rangle
      =\langle\psi,\langle Q_3\rangle^{-1}A\varphi\rangle
      =\langle\langle Q_3\rangle^{-1}\psi,A\varphi\rangle\:\!,
    \end{equation}
    where $\langle\:\!\cdot\:\!,\:\!\cdot\:\!\rangle$ denotes the duality between $\dom(H_0)$
    and $\dom(H_0)^*$. Since $\langle Q_3\rangle^{-1}$ is a homeomorphism from $\dom(H_0)^*$
    to the domain of $\langle Q_3\rangle$ in $\dom(H_0)^*$, it follows from (\ref{cidessus})
    that the domain of $\langle Q_3\rangle$ in $\dom(H_0)^*$ is included in the domain of $A$
    in $\dom(H_0)^*$ (the adjoint of the operator $A$ in $\dom(H_0)$ is equal
    to the operator $-A$ in $\dom(H_0)^*$).

    (iv) The inequality
    $r\:\!\|(\langle Q_3\rangle+ir)^{-1}\|_{\B\(\dom(H_0)^*\)}\leq\textrm{Const.}$ for all
    $r>0$ is obtained from relation (\ref{Naudts}), given in the proof of Lemma \ref{Hetf},
    with $f(Q_3)=(\langle Q_3\rangle +ir)^{-1}$.

    (v) Assume that $V$ is a long-range (scalar) potential. Then the following equality holds
    in form sense on $\D$\:\!:
    \begin{equation}\label{ciavant}
      2\[iV,A\]=-Q_3(\partial_3 V)H_0^{-1}-H_0^{-1}Q_3(\partial_3 V)
      +\[iV,H_0^{-1}\]Q_3 P_3 + P_3Q_3\[iV,H_0^{-1}\], 
    \end{equation}
    with $\[iV,H_0^{-1}\]=\sum_{j=1}^3H_0^{-1}\alpha_j(\partial_jV)H_0^{-1}$. Using Lemma
    \ref{Hetf}.(a), one gets that the last two terms in (\ref{ciavant}) are equal in form
    sense on $\D$ to
    \begin{equation*}
      2\re\sum_{j=1}^3H_0^{-1}\alpha_jQ_3(\partial_jV)P_3H_0^{-1}
      -2\im\sum_{j=1}^3 H_0^{-1}\alpha_j(\partial_jV)H_0^{-1}\alpha_3P_3H_0^{-1}.
    \end{equation*}
    It follows that $\[iV,A\]$, defined in form sense on $\D$, extends continuously to an
    operator in $\B(\H)$. Now let $\vartheta$ be as in Definition \ref{defdespots}. Then
    a direct calculation using the explicit form of $[iV,A]$ obtained above implies that
    \begin{equation*}
      \Big\|\vartheta\Big(\frac{\langle Q_3\rangle}r\Big)\[iV,A\]\Big\|
      \leq\textsc c\:\!\sum_{j=1}^3\Big\|\vartheta\Big(\frac{\langle Q_3\rangle}r\Big)
      \langle Q_3\rangle(\partial_jV)\Big\|+\frac{\textsc d}r
    \end{equation*}
    for all $r>0$ and some positive constants $\textsc c$ and $\textsc d$.
  \end{proof}
  
  As a direct a consequence, one obtains that
  
  \begin{Lemma}\label{Mourre for H}
    If \:\!$V$ satisfies the hypotheses of Theorem \ref{final thm}, then 
    $A$ is conjugate to $H$ on \:\!$\R\setminus\sigma^0_{\rm sym}$\:\!.
  \end{Lemma}

  \begin{proof}
    Proposition \ref{V regularity} implies that both $H_0$ and $H$ are of class
    $\mathscr C^{1,1}(A)$. Furthermore, the difference $(H+i)^{-1}-(H_0+i)^{-1}$ is compact
    by Proposition \ref{compactness}, and $\varrho^T_{H_0}>0$ on
    $\R\setminus\sigma_{\rm sym}^0$ due to Proposition \ref{Mourre for H_0}. Hence the
    claim follows by \cite[Thm.~7.2.9\,\&\,Prop.~7.2.6]{ABG}. 
  \end{proof}

  We can finally give the proof of Theorem \ref{final thm}.

  \begin{proof}[Proof of Theorem \ref{final thm}]
    Since $A$ is conjugate to $H$ on \:\!$\R\setminus\sigma^0_{\rm sym}$ by Lemma
    \ref{Mourre for H}, the assertions (a) and (b) follow by the abstract conjugate
    operator method~\cite[Cor.~7.2.11\,\&\,Thm.~7.4.2]{ABG}.
    
    The limiting absorption principle directly obtained via \cite[Thm.~7.4.1]{ABG} is
    expressed in terms of some interpolation space, associated with $\dom(A)$, and of its
    adjoint. Since both are not standard spaces, one may use \cite[Prop.~7.4.4]{ABG} for
    the Friedrichs couple $(\dom(\langle Q_3\rangle),\H)$ to get the statement (c). In
    order to verify the hypotheses of that proposition, one has to check that for each
    $z\in\C\setminus\sigma(H)$ the inclusion
    $(H-z)^{-1}\dom(\langle Q_3\rangle)\subset\dom(A)$ holds. However, since
    $\dom(\langle Q_3\rangle)$ is included in $\dom(A)$ by Proposition \ref{sur A}, it is
    sufficient to prove that for each $z\in\C\setminus\sigma(H)$ the operator $(H-z)^{-1}$
    leaves $\dom(\langle Q_3\rangle)$ invariant. Since $\dom(H)=\dom(H_0)$ is left
    dinvariant by the group $\{e^{it\langle Q_3\rangle}\}_{t\in\R}$ (see Proposition
    \ref{V regularity}\,(i)) one easily gets from \cite[Thm.~6.3.4.(a)]{ABG} that $H$ is
    of class $C^1(\langle Q_3\rangle)$, which implies the required invariance of
    $\dom(\langle Q_3 \rangle)$ \cite[Thm.~6.2.10.(b)]{ABG}.
  \end{proof}

  \section*{Appendix}
  
  \begin{proof}[Proof of Lemma \ref{P_3H_0^{-1}}]
    (a) Let $\varphi$, $\psi$ be in $\D$. Using the transformation (\ref{F unitaire}),
    one gets
    \begin{equation*}
      \<H_0^{-n}\varphi,Q_3\psi\>=\int_\R\<H_0(\xi)^{-n}(\F\varphi)(\xi),
      (i\partial_\xi\F\psi)(\xi)\>_{\H_{\scriptscriptstyle 12}}\de\xi\:\!.
    \end{equation*}
    Now the map 
    $\R\owns\xi\mapsto H_0(\xi)^{-n}\in\B(\H_{\scriptscriptstyle 12})$ is norm
    differentiable with its derivative equal to
    $-\sum_{j=1}^nH_0(\xi)^{-j}\alpha_3H_0(\xi)^{j-n-1}$.
    Hence
    $\{\partial_\xi[H_0(\xi)^{-n}(\F\varphi)(\xi)]\}_{\xi\in\R}$ belongs to
    $\int_\R^\oplus\H_{\scriptscriptstyle 12}\,\de\xi$. Thus one can perform an
    integration by parts (with vanishing boundary contributions) and obtain
    \begin{equation*}
      \<H_0^{-n}\varphi,Q_3\psi\>=\int_\R\<i\partial_\xi[H_0(\xi)^{-n}(\F\varphi)(\xi)],
      (\F\psi)(\xi)\>_{\H_{\scriptscriptstyle 12}}\de\xi\:\!.
    \end{equation*}
    It follows that
    $\left|\<H_0^{-n}\varphi,Q_3\psi\>\right|\leq\textrm{Const.}\:\!\|\psi\|$ for all
    $\psi\in\D$. Since $Q_3$ is essentially self-adjoint on $\D$, this implies that
    $H_0^{-n}\varphi$ belongs to $\dom(Q_3)$.
    
    (b) The boundedness of $P_3H_0^{-1}$ is a consequence of the estimate
    \begin{equation*}
      \esssup_{\xi\in\R}\|\xi H_0(\xi)^{-1}\|_{\B(\H_{\scriptscriptstyle 12})}
      =\esssup_{\xi\in\R}\left\|\frac{|\xi|}{[H_0(0)^2+
      \xi^2]^{1/2}}\right\|_{\B(\H_{\scriptscriptstyle 12})}<\infty
    \end{equation*}
    and of the direct integral formalism \cite[Prop.~3.6\;\&\;3.7]{Cho}. The
    remaining assertions follow by standard arguments.
  \end{proof}

  \begin{proof}[Proof of Lemma \ref{Hetf}]
    (a) One first observes that the following equality holds on $\D$\:\!:
    \begin{equation}\label{Naudts}
      iH_0^{-1}f(Q_3)H_0=-H_0^{-1}\alpha_3f'(Q_3)+if(Q_3)\:\!.
    \end{equation}
    Now, for $\varphi, \psi\in\D$ and $\eta \in H_0^{-n}\D$, one has
    \begin{align*}
      &\langle\varphi,iH_0^{-1}f(Q_3)\eta\rangle
      -\langle\varphi,if(Q_3)H_0^{-1}\eta\rangle\\
      &=\langle\varphi,iH_0^{-1}f(Q_3)H_0\psi\rangle
      +\langle\varphi,i H_0^{-1}f(Q_3)(\eta-H_0\psi)\rangle
      -\langle\bar f(Q_3)\varphi,iH_0^{-1}\eta\rangle\\
      &=-~\langle\varphi,H_0^{-1}\alpha_3f'(Q_3)H_0^{-1}\eta\rangle
      -\langle \varphi,H_0^{-1}\alpha_3 f'(Q_3)H_0^{-1}
      (H_0\psi-\eta)\rangle\\
      &\qquad+\langle\bar f(Q_3)\varphi,iH_0^{-1}(H_0\psi-\eta)\rangle
      +\langle\bar f(Q_3)H_0^{-1}\varphi,i(\eta-H_0\psi)\rangle\:\!,
    \end{align*}
    where we have used (\ref{Naudts}) in the last equality for the term 
    $\langle\varphi,iH_0^{-1}f(Q_3)H_0\psi\rangle$. Hence there exists
    a constant $\textsc c$ (depending on $\varphi$) such that
    \begin{equation*}
      |\langle\varphi,iH_0^{-1}f(Q_3)\eta\rangle
      -\langle\varphi,if(Q_3)H_0^{-1}\eta\rangle 
      +\langle \varphi,H_0^{-1}\alpha_3f'(Q_3)H_0^{-1}\eta\rangle|
      \leq\textsc c\;\!\|\eta-H_0\psi\|\:\!.
    \end{equation*}
    Then the statement is a direct consequence of the density of $H_0\D$
    and $\D$ in $\H$.

    (b) This is a simple corollary of the point (a).
  \end{proof}
  
  \section*{Acknowledgements}
  
  We thank M. M\v antoiu for having suggested to us the present study. We are
  also grateful to W. Amrein, H. D. Cornean and B. Thaller for their helpful
  remarks. This work was partially supported by the Swiss National Science
  Foundation.
  

\end{document}